\documentclass[showpacs,twocolumn,aps,pra,superscriptaddress]{revtex4}

\usepackage{graphicx}

\begin{document}

\title{Controlling excess noise \\ in fiber optics continuous variables
quantum key distribution}

\author{J\'er\^ome Lodewyck}
\affiliation{Thales Research and Technologies, RD 128\\
 91767 Palaiseau CEDEX France}
\affiliation{Laboratoire Charles Fabry de l'Institut d'Optique, Campus
Universitaire, b\^at 503\\ 91403 Orsay CEDEX France}
\author{Thierry Debuisschert}
\affiliation{Thales Research and Technologies, RD 128\\
 91767 Palaiseau CEDEX France}
\author{Rosa Tualle-Brouri}
\affiliation{Laboratoire Charles Fabry de l'Institut d'Optique, Campus
Universitaire, b\^at 503\\ 91403 Orsay CEDEX France}
\author{Philippe Grangier}
\affiliation{Laboratoire Charles Fabry de l'Institut d'Optique, Campus
Universitaire, b\^at 503\\ 91403 Orsay CEDEX France}

\begin{abstract}
We describe a continuous variables coherent states quantum key distribution
system working at 1550 nm, and entirely made of standard fiber optics and
telecom components, such as integrated-optics modulators, couplers and fast
InGaAs photodiodes. The setup is composed of an emitter randomly modulating a
coherent state in the complex plane with a doubly Gaussian distribution, and a
receiver based on a shot noise limited time-resolved homodyne detector. By using
a reverse reconciliation protocol, the device can transfer a raw key rate up to
1~Mb/s, with a proven security against Gaussian or non-Gaussian attacks. The
dependence of the secret information rate of the present fiber set-up is studied
as a function of the line transmission and excess noise.
\end{abstract}



\maketitle


Quantum key distribution (QKD) is a cryptographic process enabling two distant
actors -- Alice and Bob -- to share a common secret key, unknown to a potential
eavesdropper -- Eve. Quantum laws enable Alice and Bob to detect Eve, and to
quantify the amount of information she acquired about the key, thus allowing
unconditionally secure information transfer. For this purpose Alice and Bob must
choose a proper encoding of information using non-commuting quantum channel
variables. These variables are generally the polarization or phase of single
photon pulses, requiring specifically developed components such as single photon
sources and photon counters.

In contrast, continuous variables QKD schemes typically use quadrature amplitude
of light beams as information carriers, and homodyne detection rather than
photon counting. For instance, a QKD scheme based on encoding information in the
phase and amplitude of bright coherent states\cite{fred:PRL} has been recently
demonstrated using a table-top setup at 780 nm \cite{fred:nature}. This scheme,
that we are also using here, is based upon the idea of ``reverse reconciliation"
\cite{fred:QIC03} to extract secret information from the data provided by
homodyne detection. In several recent articles, this protocol, that we will
denote as RRCS (Reverse Reconciliated Coherent State), has been proven able to
transmit secret keys for arbitrary channel transmission, and  to be secure
against non-Gaussian \cite{fred:entropic} and collective
\cite{fred:collective,navascues:collective} attacks.

The security proof of RRCS presented in  \cite{fred:entropic}, makes use of
entropic Heisenberg inequalities to set an upper bound on Eve and Bob's Shannon
mutual information $I_{BE}$ about the key. This bound is computable from the
transmission signal to noise ratio (SNR). A secret key can possibly be extracted
by error correction and privacy amplification based on Bob's copy of the key if
$$
	\Delta I = I_{AB} - I_{BE} > 0.
$$
This inequality can be satisfied in principle for any channel transmission, as
it is also the case for photon-counting QKD. This means that vacuum noise, which
is the continuous-variable equivalent to the photon losses encountered in photon
counting quantum cryptography, does not limit by itself the range of QKD. The
real limitation comes from errors in photon counting QKD
\cite{brassard:limitation}, and it is associated with {\bf excess noise} in the
case of continuous variables. By definition, excess noise is the noise {\bf
above} the vacuum noise level associated with channel losses,  and it is a major
issue in continuous variables QKD, as pointed out in various recent papers
\cite{fred:QIC03,hirano:excess}.

In particular, it has been shown in \cite{fred:QIC03} that when the excess noise
(referred to the channel input) reaches two times the shot-noise level, Eve can
perform an intercept-resend attack on the channel and thus no secure key
can be transmitted. As another illustration of the importance of excess noise,
it has been pointed out in \cite{hirano:decline} that the presence of excess
noise severely weakens protocols that use post-selection to extract bits from
the correlated Gaussian distributions shared by Alice and Bob
\cite{leuchs:post,ralph:post}.

A specific feature of the RRCS protocol is to use full Gaussian distributions
for data transmission between Alice and Bob, enforcing that the optimal attack
by Eve is also Gaussian \cite{fred:entropic}. In case Eve would like to try a
less efficient non-Gaussian attack, the distribution received by Bob may not be
Gaussian anymore, but the information acquired by Eve remains bounded by the
variance of the noise measured by Alice and Bob. This is a very convenient
feature, which requires however to extract very efficiently the bits from the
correlated Gaussian data. The combination of these features -- Gaussian
modulation and the resulting possibility to evaluate analytically the tolerance
to excess noise -- warrants that the secret bit rate can be evaluated simply
from real transmission data.


It is also worth noticing that the use of quadrature modulation and homodyne
detection in the RRCS protocol is well suited for telecom application, because
it can be implemented with off the shelves fast and efficient telecom
components, and use existing single mode telecom fibers. For this reason, it
features a very high nominal secret key rate. As it will be discussed below, the
target distance range for this kind of setup is several tens of kilometers,
limited by the efficiency of classical bit error correction algorithms.

In this article, we describe an all-fiber optics continuous variables RRCS-QKD
set-up, and give an explicit evaluation of the available secret bit rate
obtained by measuring the signal to noise degradation for various values of the
line transmission, in presence of technical excess noise. These results show
clearly the importance of evaluating and controlling excess noise in continuous
variables QKD setups for a proper determination of the secret key rate.

\begin{figure}
	\center{\includegraphics[width=\columnwidth]{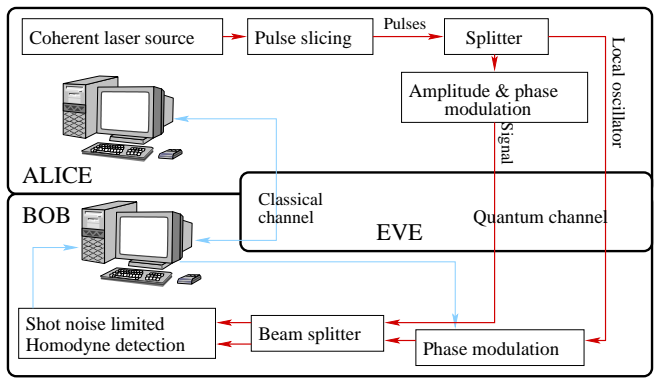}}
	\caption{(Color online) Experiment layout. Alice sends both modulated signal
	and phase reference to Bob. A random quadrature is measured by a
	time-resolved, shot noise limited homodyne detection.}
	\label{fig:experiment}
\end{figure}

The experimental scheme is composed of two independent modules. Alice, the
sender, has to randomly displace coherent states in the complex Fresnel plane.
Bob, the receiver, has to measure a random quadrature of the incoming signal,
with a shot-noise limited pulsed homodyne detector (see figure
\ref{fig:experiment}).

Alice first generates 100 ns wide pulses from a continuous wave DFB laser diode
emitting at 1550 nm with an integrated electro-optics modulator, driven by a
pulse generator. These pulses are split into a strong phase reference (or local
oscillator) of $10^8$ photons per pulse, and a weak signal (typically 100
photons). Each signal pulse is a well-defined coherent quantum state of the
light representing a channel symbol. To compensate for various losses in the
phase reference optical path, the production of these pulses requires a laser
power of about 100 mW at the diode output. Alice's setup is entirely made of
polarization maintaining fibers in order to avoid polarization drifts at the
modulators input, and relative polarization drifts between signal and local
oscillator.

This signal is continuously modulated in phase and amplitude with
computer-driven electro-optics amplitude and phase modulators, in order to place
the coherent states in the complex plane. For our continuous variables QKD
protocol, the required modulation is a two-dimensional Gaussian distribution
centered on zero, with a customizable variance. Due to modulators dynamics, the
modulation is truncated to four standard deviations, thus resulting in an error
of 0.3\% on variances estimations. For a modulation variance of 40 photons per
pulse, the modulation inaccuracy is typically less than 4 percents relative to
the shot noise variance, at rates up to 1~MHz. This 4\% inaccuracy is equivalent
to an excess noise (see below). It is mainly due to non-perfect static and
dynamic modulator voltage settings, as well as voltage rising time for large
modulation steps.


In the present implementation, the signal and phase reference are sent to Bob by
using two separate fibers with a length of a few meters, properly isolated from
external perturbations. In this configuration, we can simulate channel losses by
varying Alice's modulation variance $V_A$. The reference level (that defines
unity gain) is set to $V_A = 40 V_0$, where the shot noise variance $V_0$ will
be used as a reference for all noise levels in the following.

While this setup is suitable for our noise analysis, it is not optimized for
field QKD, because over long distances two different single mode fibers will
see large relative polarization and phase drifts. To get rid of these
perturbations, the signal and local oscillator should be sent in the same fiber
with a time delay. We have made preliminary tests for such a time-multiplexing
into a single fiber, using also an active polarization controller to avoid
unwanted polarization drifts. The results are promising, but a full key
distribution has yet to be performed in this configuration.


Bob's setup is composed of a shot noise limited time resolved homodyne
detection. Weak signal pulses interfere with the phase reference,
and light intensity in both output arms is measured with matched fast InGaAs
photodiodes (10 GHz, 80\% efficiency). The signal quadrature is then
obtained by subtraction of the two photocurrents, amplified with a low noise
charge amplifier \cite{hansen} followed by a constant gain amplifying stage.
Electronic noise from this amplification chain is 20 dB below the shot noise. A
time domain homodyne detection requires a precise balancing of the two arms with
an accuracy better than $10^{-4}$ so that the residual unbalance does not
saturate amplifiers. This is achieved with mechanical fiber optics variable
attenuators, which introduce small losses by bending the fiber. Such a balancing
is very stable on time scales of several hours. Bob can select a desired
quadrature at 1~MHz rate with a phase modulator placed in the reference optical
path. The present overall efficiency of the homodyne detector is about 60\%.
Photodiodes account for half of the losses, the remaining losses are due to
fiber connectors. Other components (coupler, variable attenuator) have very low
intrinsic insertion losses (typically 0.05 dB). To enhance information rate,
losses within the homodyne detection can be reduced by splicing fibers.


\begin{figure}
	\center{\includegraphics[width=\columnwidth]{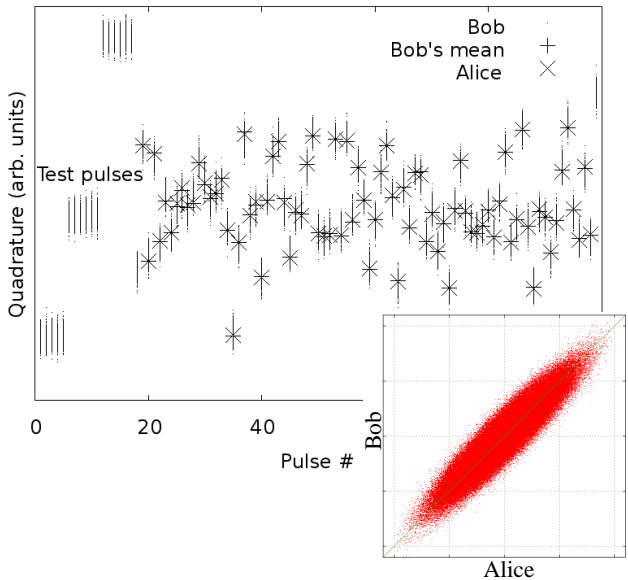}}
	\caption{(Color online) Random modulation. Alice's random modulation is
repeated every 100 pulses in order to observe the shot noise and modulation
imperfections. Points are Bob's measurement, $\times$ crosses are Alice's
modulation and $+$ crosses are Bob's measurement average. The difference between
$+$ and $\times$ crosses represents the technical noise due to modulation
imperfections, typically 4\% of the shot noise variance $V_0$ for a modulation
variance of 40 $V_0$. These 100 pulses contain 20 test pulses (left part) and 80
useful modulation pulses. With truly random modulation we obtain correlation
between Alice and Bob (bottom-right), the width of the data set showing the
total noise level referred to the input.}
	\label{fig:random}
\end{figure}

Alice and Bob are computer-interfaced by a synchronous automatic data processing
software. In order to fully implement a QKD scheme, we designed a communication
protocol that can synchronize Alice and Bob and provide for channel parameters
(gain, excess noise and relative phase). The communication is split into
independent blocks. The block size -- typically 50000 pulses -- is adjusted so
that we can assume transmission parameters are constant over a block, while
being large enough to make statistical tests. A block is composed of smaller
(100 pulses long) structures, containing 80 useful modulation pulses and 20 test
pulses (figure \ref{fig:random}), forming a software detectable pattern. These
test pulses consist of maximal amplitude and phase modulated coherent states.
From these pulses, one can synchronize Alice and Bob, determine the mean signal,
the relative phase between signal and local oscillator, and the phase and
intensity noise. The properties of the test pulses are chosen so that all these
parameters can be independently determined. By averaging test pulses over a
block, we can get rid of the shot noise and obtain accurate channel parameters
determination. We note that the test pulses might be used or manipulated by the
eavesdropper, and therefore all calibrations must also be double-checked by
statistics over a randomly chosen revealed sample from the Gaussian data set.


\begin{figure}
	\center{\includegraphics[width=\columnwidth]{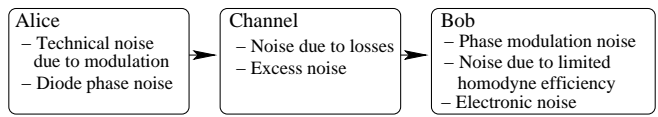}}
	\caption{Nature of noise sources found in the QKD device. Technical, phase
	and electronic noises account for the total excess noise.}
	\label{fig:noise}
\end{figure}

The setup described so far produces correlated Gaussian-distributed continuous
variables at a 1~MHz rate. In order to evaluate the raw key rate from these
correlations, we need to review noises sources appearing in the protocol (figure
\ref{fig:noise}). As said above, the decrease in SNR during propagation in the
channel can be split into two different terms: the vacuum noise due to line
losses, and the excess noise. In this picture, we can write the total added
noise $\chi$, referred to the input and expressed in shot noise units, as
$ \chi = (1-G)/G + \xi$,
where $G$ is the channel gain and $\xi$ is the excess noise.

Even in the absence of Eve, excess noise is introduced by technical imperfection
in our modulation system and by the laser diode phase noise. In principle, this
excess noise is not due to Eve and could be considered unknown to her. However,
the level of modulation and phase excess noise is drifting, depending on the
modulators settings, and cannot be calibrated. Therefore, it is wiser to assume
that it can be generated and controlled by Eve. Let us emphasize also that
fiber optics without repeaters or amplifiers do not generally introduce excess
noise. However, uncertainty on Bob's estimation of the output noise is
equivalent to an excess noise at the input, function of the line transmission,
potentially accessible to Eve.

The phase noise level for maximum output intensity is typically 0.2 $V_0$, but
it is as low as 0.01 $V_0$ when averaged over the Gaussian modulation. To
achieve such a low phase noise, the laser diode must be strongly attenuated
($\geq$ 80 dB) from its initial power of 100 mW, and the path difference between
interferometer arms (a few tens of centimeters) has to be small compared to the
laser coherence length. All in all, the total excess noise measured is $\xi =
0.06 \; V_0$ for a modulation variance $V_A = 40 \; V_0$, and decreases
proportionally for lower variances.

\begin{figure}
\center{\includegraphics[width=\columnwidth]{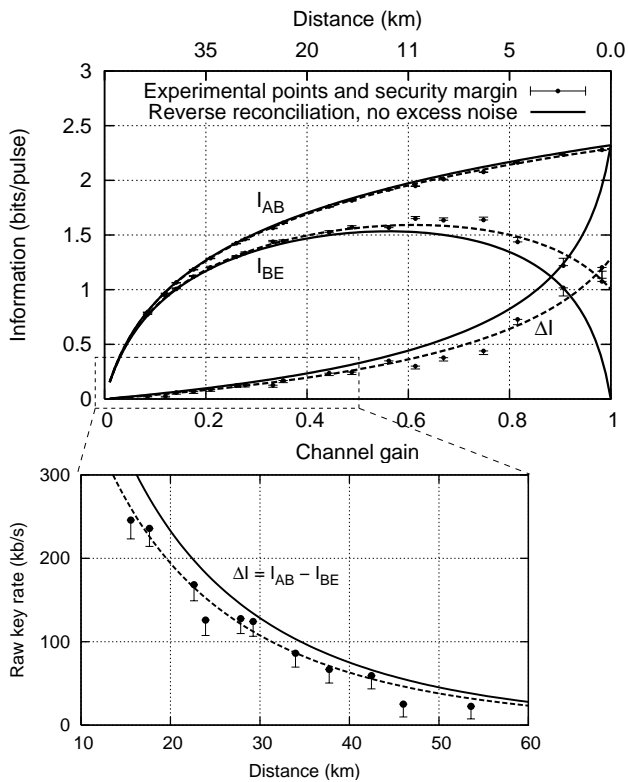}}
	\caption{Shannon mutual information per pulse shared by Alice and Bob. Solid
lines define reverse reconciliation Shannon information rates for a lossy
channel without excess noise. Because of technical noise, the experimental
(dots) and theoretical (dashed lines) information rates achievable by the
experiment are lower. The figure also shows security margins of 0.02 $V_0$ units
on Bob's output excess noise evaluation. The bottom left figure is a zoom
of the advantage of Alice on Eve in terms of available information rate before
binary processing. It is plotted as a function of the distance rather than the
channel gain. It is equivalent to the secret key rate that would be obtained
after applying perfect
($\beta = 1$) error correction and privacy amplification to our data.}
	\label{fig:information}
\end{figure}

Losses in Alice's device do not matter since the reference level is calibrated
at Alice's output. Losses of the homodyne detector, while deteriorating Bob's
SNR, can be considered unrelated to Eve, and therefore do not contribute to her
information. This approach, that considers Eve has no access to Bob's hardware,
is called ``realistic mode", by opposition to a ``paranoid mode" where Eve would
be able to exploit internal defects of Bob's setup. In any case, it is clear
that the homodyne detector efficiency $\eta$ needs to be very carefully
calibrated. In this realistic picture, we can derive mutual Shannon information
rates as a function of the channel gain $G$ and the excess noise $\xi$
\cite{fred:nature}:
\begin{eqnarray*}
	I_{AB} & = & \frac 1 2 \log_2 \frac{\eta G V_A+1+\eta G \xi}{1+\eta G \xi}\\
	I^\textrm{max}_{BE} & = &
	\frac 1 2 \log_2 \frac{\eta G V_A+1+\eta G \xi}
	{\eta/\left[1-G+G\xi+G/(V_A+1)\right] + 1 - \eta}.
\end{eqnarray*}
All the quantities appearing in these formulas are measured by Alice and Bob.
Experimental measurement of $\xi$ for different channel transmissions enables to
plot the rates $I_{AB}$ and $I_{BE}$ achieved by our setup as a function of $G$
(figure \ref{fig:information}). For this plot the homodyne efficiency $\eta$ is
0.6, the modulation variance $V_A$ is set to 40 $V_0$, and the excess noise
$\xi$ is either zero (solid lines) or 0.06 $V_0$ (dashed lines). The graph
clearly shows that $\Delta I = I_{AB} - I_{BE}$ remains positive even for low
transmission, equivalent to a 55 km propagation distance, including security
margins in excess noise evaluation.


Given this raw available secret information rate, the secret bits still have to
be extracted from the Gaussian data. Presently this is done using a ``sliced
reconciliation'' algorithm \cite{gilles:ieee}, with an efficiency which is
typically 0.7 to 0.8 of Shannon's limit in our operating conditions. Eve's
information about the key is finally erased by a standard privacy amplification
procedure. These algorithms are being interfaced with the experiment.The present
version of reconciliation algorithm implements true one-way
reconciliation based on turbo-codes \cite{gilles:turbo}, which eliminates the
need for extra assumptions when using RRCS protocols. We point out however that
if the efficiency of the algorithms is $\beta$ (with respect to Shannon's
entropy), the key rate drops to $\Delta I_\textrm{eff} = \beta  I_{AB}-I_{BE}$,
and vanishes beyond approximately 20 km. Another limitation is the speed of
reconciliation algorithms, which is currently able to process data at about
100~kHz, using an average PC. Work is underway to improve both the speed and the
efficiency of the algorithms.


As a conclusion, the setup described in this paper is functional and ready to be
tested on a field scale. The current data rate is limited by the data
acquisition and processing, rather than  by optical components which can go as
fast as 10~GHz. The homodyne detection can be extended to 100~MHz \cite{zavatta}
by using dedicated electronics for Alice and Bob (rather than personal
computers). As a consequence, rather straightforward extensions of this setup
should yield up to 100~Mb/s raw key rate in a low-loss line. The ultimate usable
secret key rate will depend on further progress in data reconciliation softwares
\cite{bloch}.
 

We would like to thank Andr\'e Villing, who made the homodyne detection
electronics. We acknowledge financial support from the European Commission
through the IST-SECOQC Integrated Project.

\end{document}